\documentclass[trackchanges]{aastex701}

\newcommand{\rh}{\ensuremath{r_{\rm h}}}

\newcommand{\mcl}{\ensuremath{M_{\rm cl}}}
\newcommand{\fbin}{\ensuremath{f_{\rm bin}}}
\newcommand{\fobin}{\ensuremath{f_{\rm Obin}}}

\newcommand{\Ms}{\ensuremath{{\rm M}_{\odot}}}

\begin{document}
\title{The velocity dispersion profile of nine  open clusters in the solar neighborhood}

\shorttitle{Velocity dispersion of open clusters}

\shortauthors{Ma et al.}

\author[0009-0005-3931-9072]{Bingqian Ma}
    \affiliation{Department of Physics, Xi'an Jiaotong-Liverpool University, 111 Ren’ai Road, Dushu Lake Science and Education Innovation District, Suzhou 215123, Jiangsu Province, P.R. China.}
    \email{mbcz31242@outlook.com}

\author[0000-0003-3389-2263]{Xiaoying Pang}
    \affiliation{Department of Physics, Xi'an Jiaotong-Liverpool University, 111 Ren’ai Road, Dushu Lake Science and Education Innovation District, Suzhou 215123, Jiangsu Province, P.R. China.}
    \email{Xiaoying.Pang@xjtlu.edu.cn}
    \affiliation{Shanghai Key Laboratory for Astrophysics, Shanghai Normal University, 
                100 Guilin Road, Shanghai 200234, P.R. China}
    
\author[0000-0002-1254-2603]{Sambaran Banerjee} 
    \affiliation{Helmholtz-Instituts f\"{u}r Strahlen-und Kernphysik, Nussallee 14-16, D-53115 Bonn, Germany; Argelander-Institut f\"{u}r Astronomie, Auf dem Hügel 71, D-53121, Bonn, Germany}
    \email{sambaran@astro.uni-bonn.de}

\author[0009-0002-4282-6073]{Pengfei Ren}
    \affiliation{Department of Physics, Xi'an Jiaotong-Liverpool University, 111 Ren’ai Road, Dushu Lake Science and Education Innovation District, Suzhou 215123, Jiangsu Province, P.R. China.}
    \affiliation{Department of Industrial and Systems Engineering, Georgia Tech Shenzhen Institute, Tianjin University, Block B, Building 6, Shenzhen International Innovation Valley, Dashi 1st Road, Xili Street, Shenzhen 518055, Guangdong Province, P.R. China.}
    \email{Pengfei.Ren20@alumni.xjtlu.edu.cn}

\author[0000-0002-1805-0570]{M.B.N. Kouwenhoven}
     \affiliation{Department of Physics, Xi'an Jiaotong-Liverpool University, 111 Ren’ai Road, Dushu Lake Science and Education Innovation District, Suzhou 215123, Jiangsu Province, P.R. China.}
    \email{T.KOUWENHOVEN@xjtlu.edu.cn}

\date{Received August 10, 2025; Revised September 29, 2025; Accepted October 22, 2025 \\ aaSubmitted to ApJ}

\begin{abstract} 
    We analyze the velocity dispersion profiles of nine open clusters in the solar neighborhood using kinematic data from Gaia DR\,3, aiming to identify potential dynamical signatures of stellar-mass black holes through a comparison of theoretical and observed dispersion profiles. The selected clusters include LP2373\,gp4, NGC\,1980, NGC\,2451A, NGC\,2516, NGC\,3532, NGC\,6475, UBC\,7, Praesepe, and Pleiades. We refine the center positions of the clusters with the \texttt{Meanshift} algorithm. Using the Markov Chain Monte Carlo method, we calculate the velocity dispersion for each cluster and construct one-dimensional velocity dispersion profiles. NGC\,2516, NGC\,3532, and NGC\,6475 show potential central cusps in their radial velocity dispersion profiles, which may indicate the presence of stellar-mass black holes. LP2373\,gp4, NGC\,6475, and Praesepe all display a negative correlation between velocity dispersion and stellar mass, indicating these clusters are approaching energy equipartition or expanding. NGC\,2516 and NGC\,3532 exhibit a positive dependence between velocity dispersion and stellar mass, which may be attributed to the preferential ejection of massive stars following dynamical interactions involving binaries or black holes. These two clusters are the only two that are dynamical not relaxed and are closest to virial equilibrium. We compare the observations with $N$-body simulations of star clusters. A comparison of observed and simulated velocity dispersion profiles reveals that NGC\,2516 and NGC\,3532 exhibit lower proper motion dispersions than model clusters. Better agreement with the observed profiles is achieved for model clusters with larger ages. This suggests that the observed clusters may have undergone rapid dynamical evolution. Our results suggest that NGC\,2516 and NGC\,3532 may host at least two stellar-mass black holes each. 
    
\end{abstract}

\section{Introduction}\label{sec:intro}

The gravitational collapse of giant molecular clouds and subsequent fragmentation result in the formation of stars. Stellar winds, ionizing radiation, and radiation pressure drive the rapid expulsion of the remaining gas (\citealt{krause_physics_2020}; \citealt{Lewis2023}). This rapid gas loss results in cluster expansion and potential dissolution via violent relaxation \citep{Ishchenko2025}. Under certain conditions, the process may also result in the formation of open clusters (OCs): gravitationally bound groups of stars containing tens to thousands of stars, that form from the same molecular cloud.

The evolution of OCs depends on physical conditions, such as the star formation efficiency (SFE), and the cluster's orbit in the galaxy. OCs with a low SFE are more likely to disintegrate after gas dissipation, while OCs with a high SFE have a greater chance of survival (\citealt{baumgardt2007}; \citealt{smith2011}; \citealt{shukirgaliyev2017}). The typical disruption time of an OC is 200\,Myr, and merely 3\% of the known OCs are older than 1\,Gyr \citep{pang2018}.
OCs are characterized by stellar members with similar distances, ages, and metallicities. OCs at different ages provide snapshots of stellar evolution over time \citep{Zerjal2023}. OCs are less dense, less massive (ranging from a few hundred to a few thousand solar masses), and younger (from a few Myr to several Gyr) than globular clusters (GCs), and follow Galactic orbits typical of the $\alpha$-poor, metal-rich disc from which they originate \citep{Cantat-Gaudin2024}. These physical properties provide a basis for studying the dynamical processes on the Galactic disk that control the evolution of OCs.

Clusters evolve dynamically through a number of processes, including two-body relaxation. Two primary manifestations of two-body relaxation in star clusters are mass segregation and energy equipartition (\citealt{McNamara1986}; \citealt{heggie2003}). 
Mass segregation is caused by more massive stars transferring kinetic energy to less massive stars in order to achieve energy equipartition, leading the former to migrate toward the cluster's center, while the less massive stars move outwards \citep{churchwell_spitzer_2009}. 
This process increases the mass concentration in the central region.
The exchange of kinetic energy between stars of different masses drives a cluster toward energy equipartition. 

However, the study by \citet{Trenti2013} suggest that clusters do not achieve full energy equipartition over their lifetimes. In OCs, the low stellar density leads to longer relaxation timescales, resulting in a lower degree of mass segregation and energy equipartition compared to GCs. Furthermore, since OCs are typically much younger than GCs, they are less dynamically evolved and exhibit weaker mass segregation. The degree of energy equipartition in a cluster can be inferred from the velocity dispersion of stars with different masses.

The velocity dispersion is a powerful observational tool for analyzing the internal dynamics of clusters, reflecting the dynamical equilibrium between cluster members. Studies of velocity dispersion in clusters reveal their dynamical state and the complex interplay of internal and external forces.
The degree of energy equipartition in a cluster is traced by the velocity dispersion of stars with different masses within the same region. As low-mass and high-mass stars exchange kinetic energy, the velocity dispersion of low-mass stars increases, while the velocity dispersion of high-mass stars decreases \citep{Aros2023}. As low-mass stars tend to migrate outward, the velocity dispersion of stars close to the cluster edge can be increased. In the full energy equipartition state, the velocity dispersion decreases with higher masses (see e.g., \citealt{Trenti2013}; \citealt{bianchini_effect_2016}).
In loosely-bound OCs, where gravitational interactions between stars are weaker, the stellar velocity dispersion is particularly sensitive to external influences. The velocity dispersion on the outskirts of these clusters is especially susceptible to perturbations from external tidal fields, often causing an increase in this velocity dispersion \citep{McNamara1986}.

In the centers of star clusters, the presence of intermediate-mass black holes (IMBHs) and stellar-mass black holes inhibits the development of energy equipartition \citep{Aros2023}.  The velocity dispersion in the central region becomes higher due to the presence of BHs, and therefore its dependence on stellar mass weakens. BHs tend to concentrate near the centers of clusters and tend to expel massive stars, inhibiting mass segregation.  Consequently, the velocity dispersion in the central region  instead exhibits a local cusp (\citealt{torniamenti_stellar-mass_2023}; \citealt{Gottgens2021}).

In their study of the velocity dispersion profile of the M\,80, \citep{Gottgens2021} suggested the presence of an IMBH. A direct evidence of an IMBH with mass near 50,000~$\Ms$ in $\omega$~Centauri was discovered by \citet{maximilian2024}. \citet{torniamenti_stellar-mass_2023} found evidence for 2–3 stellar-mass BHs in the Hyades cluster, based on a comparison of the velocity dispersion profile with that of $N$-body simulations.
However, previous studies did not consider the effect of binary systems on the velocity dispersion profile. In this paper, we investigate the dynamical signatures of BHs in open clusters by analyzing velocity dispersion profiles, including the effects of binary motions on velocity dispersion. We carry out a detailed comparison between theoretical and observed velocity dispersion profiles and aim to identify potential stellar-mass BHs in OCs.

Directly calculating the velocity dispersion using the standard deviation ignores observational errors, and is also sensitive to outliers. The orbital motions of binary stars, as well as the potential escaping members, artificially increase the measured velocity dispersion \citep[e.g,][]{kouwenhoven2008, torniamenti_stellar-mass_2023}.
To address this issue, previous studies have introduced models to improve accuracy. For example, \citet{Baumgardt2018} introduced a Gaussian model to account for the velocity dispersion inflation caused by binaries, enhancing measurement accuracy. However, this method does not eliminate contamination from field stars and remains sensitive to outliers.
To achieve more precise velocity dispersion measurements, our current study adopts the approach developed by \citet{pang2018}. This method uses two Gaussian models to separately account for field stars and binaries, providing more accurate velocity dispersion estimates.
Furthermore, in our study we obtain proper motion (PM) and original radial velocity (RV) data for cluster stars from the Gaia DR\,3 catalog \citep{Gaia2023}. These more accurate datasets enhances the reliability of our velocity dispersion measurements.

This paper is organized as follows. In Section~\ref{sec:gaia} we introduce the dataset and the membership of the star clusters, followed by an explanation of \texttt{StarGO}, the algorithm used to determine cluster membership. In Section~\ref{sec:veldisp}, we measure the 1D velocity dispersion profiles of the clusters, and analyze the dependence of velocity dispersion on stellar mass for each cluster. The calibration of the cluster centers is performed using the \texttt{Meanshift} algorithm in Section~\ref{sec:cc}. In Section~\ref{sec:nbody}, we describe  the $N$-body simulations of the open clusters. In Section~\ref{sec:comparison}, we describe and compare the simulated results with the observed results. In Section~\ref{sec:dis}, we discuss our results. Finally, we summarize our conclusions in Section~\ref{sec:sum}. 


\section{Open clusters Membership}\label{sec:gaia}

The analysis is carried out for intermediate-mass star clusters with sufficient kinematic data from Gaia DR\,3 \citep{Gaia2023}, which ensures that a reliable velocity dispersion analysis can be carried out in this paper. We select nine clusters in the solar neighborhood, LP2373\,gp4, NGC\,1980, NGC\,2451A, NGC\,2516, NGC\,3532, NGC\,6475, UBC\,7, Praesepe and Pleiades. In this study, we use the Pang star cluster catalog \citep{pang2024} based on Gaia DR\,3.

The pre-processing before member selection follows our previous study. We filter out stars with parallax and photometric relative uncertainties exceeding 10\%, and apply additional astrometric cuts described in appendix~C of \cite{lindegren2018}. We apply a PM cut for each cluster region based on a 2D PM density map. We use the average spatial coordinates and proper motions of clusters from \citet{liu2019} and \citet{cantat2020}. Stars with angular separation below the Gaia angular resolution limit \citep[0.6 arcsecond,][]{lindegren_gaia_2021} are treated as single stars from the Gaia pipeline, and are therefore deemed unresolved binaries \citep{pang2023}. Though our nine clusters are all located within 500\,pc, the majority of the binary systems in each cluster still remain unresolved.

The selection of star cluster members is executed using the machine learning algorithm \texttt{StarGO} \citep{yuan2018}, based on the Self-Organizing-Map (SOM) algorithm. In short, the SOM algorithm starts by constructing a 2D neural network. Each neuron is assigned equal dimensional random weight vectors based on the input parameters. The inputs in this study are five-dimensional (5D) parameters ($X$, $Y$, $Z$, \(\mu_{\alpha}\cos\delta\), and \(\mu_{\delta}\)), where $X$, $Y$, $Z$ are Heliocentric coordinates, and \(\mu_{\alpha}\cos\delta\) and \(\mu_{\delta}\) are the components of the proper motion from Gaia DR\,3. Depending on the number of input stars, we use a 100×100 or 150×150 neuron network. During each iteration, the weight vector of each neuron is updated so that it is closer to the input vector of an observed star. For the weight vectors in all neurons to converge, we set the number of iterations to 400. At the end of the iterations, each star has its corresponding neuron whose weight vector is the closest to the 5D parameters of that star.

In a 2D neuron network, the value of the difference in weight vectors between adjacent neurons, \texttt{u}, is used to cluster similar and nearby neurons into groups. The final cluster members are the neurons corresponding to stars below a specific threshold, which is chosen to satisfy a field star contamination rate of 5\%. The estimation of the field star contamination rate is computed using stars from the mock Gaia DR\,3 catalog \citep{rybizki2020}.

\begin{samepage}
\startlongtable
\begin{deluxetable*}{L L RRRRR CCC LL R}
\tablecaption{Properties of the 9 open clusters in this study. 
\label{tab:fdim}}
	\tabletypesize{\scriptsize}
	\tablehead{
            \colhead{Cluster}    & 
            \colhead{Age}       &
            \colhead{R.A.}       &
		\colhead{Decl.}      & 
            \colhead{$X_c$}        &
            \colhead{$Y_c$}        &
            \colhead{$Z_c$}        &                     
            \colhead{$\sigma_{\mu_{\alpha}\cos\delta}$}    &
            \colhead{$\sigma_{\mu_{\delta}}$} &
            \colhead{$\sigma_{\rm RV}$}      &
            \colhead{$N_{\rm RV}$}      &
            \colhead{$M_{\rm total}$}      &
            \colhead{$M_{\rm limit}$}   
		\\
		\colhead{}              & 
            \colhead{(Myr)}       &
		\colhead{(deg)}         & 
		\colhead{(deg)}         &
		\colhead{(pc)}          &
  	    \colhead{(pc)}          &
		\colhead{(pc)}         &
            \colhead{(km\,s$^{-1}$)}      &
            \colhead{(km\,s$^{-1}$)}      &
            \colhead{(km\,s$^{-1}$)}     &
            \colhead{}     &
            \colhead{$(\Ms)$}     &
            \colhead{$(\Ms)$}   
		\\
            \cline{2-3}
	    \cline{4-6}
            \cline{7-13}
	    \colhead{(1)}   & 
	    \colhead{(2)}   & \colhead{(3)} & 
	    \colhead{(4)}   & \colhead{(5)} &
	    \colhead{(6)}   & \colhead{(7)} & 
	    \colhead{(8)}   & \colhead{(9)} &
            \colhead{(10)}   & \colhead{(11)} &
             \colhead{(12)} & \colhead{(13)}
		}
	\startdata

\mathrm{LP\ 2373\ gp4} & 6.2 & 84.2 & -2.0 & -310.7 & -152.2 & -109.1 &  0.27\pm0.01 & $0.25_{\text{-0.01}}^{\text{+0.02}}$ & $19.37_{\text{-1.58}}^{\text{+1.66}}$ & 139 & 295.6 & 0.25\\

\mathrm{NGC\ 1980} & 5.9 & 83.8 & -5.8 & -314.8 & -177.4 & -128.7 &  0.31\pm0.01 & 0.37\pm0.01 & 18.87\pm1.22 & 329 & 757.2 & 0.26\\

\mathrm{NGC\ 2451A} & 58 & 115.7 & -38.2 & -57.8 & -182.1 & -24.3 &  0.26\pm0.04 & 0.53\pm0.02 & $1.38_{\text{-0.58}}^{\text{+0.64}}$ & 91 & 178.9 & 0.23\\

\mathrm{NGC\ 2516} & 123 & 119.5 & -60.8 & 26.5 & -394.0 & -112.4 &  0.51\pm0.01 & 0.43\pm0.01 & $1.88_{\text{-0.16}}^{\text{+0.17}}$ & 831 & 1984.8 & 0.36\\

\mathrm{NGC\ 3532} & 398.1 & 166.3 & -58.7 & 159.9 & -450.6 & 11.5 &  0.37\pm0.01 & 0.39\pm0.01 & 1.36\pm0.14 & 1051
& 2228 & 0.38\\

\mathrm{NGC\ 6475} & 186 & 268.4 & -34.8 & 278.0 & -20.2 & -21.9 &  0.49\pm0.01 & 0.502\pm0.01 & 1.55\pm0.15 & 469 & 1023.1 & 0.36\\

\mathrm{Pleiades} & 125 & 56.6 & 24.1 & -121.1 & 29.1 & -54.3 & 1.29\pm0.03 & 1.8\pm0.04 & $1.67_{\text{-0.15}}^{\text{+0.17}}$ & 571 
& 740.6 & 0.28\\

\mathrm{Praesepe} & 700 & 130.0 & 19.6 & -140.5 & -68.4 & 99.0 & $1.32_{\text{-0.04}}^{\text{+0.11}}$ & 0.96\pm0.02 & 0.97\pm0.07 & 428 & 601.5 & 0.30\\

\mathrm{UBC\ 7} & 40 & 106.9 & -37.6 & -97.9 & -252.6 & -63.7 & 0.32\pm0.2 & 0.35\pm0.02 & $0.33_{\text{-0.23}}^{\text{+0.4}}$ & 91 & 191.6 & 0.21
    \enddata
	\begin{tablecomments}{
        Age is the cluster age. R.A. and Decl. are the right ascension and declination center. $X_c$, $Y_c$ and $Z_c$ indicate the cluster centers in Cartesian coordinates. $\sigma_{\rm RV}$ is the RV dispersion. $\sigma_{\mu_{\alpha}\cos\delta}$ and $\sigma_{\mu_{\delta}}$ are the dispersions of the R.A. and Decl. components of the PMs. $M_{\rm limit}$ is the lowest mass of the observed cluster members. $M_{\rm total}$ is the total mass of the observed cluster. $N_{\rm RV}$ is the members with accurate Gaia DR\,3 RV data. The ages, $M_{\rm limit}$ and $M_{\rm total}$ are taken from \citet{pang2024}.
	    }
    \end{tablecomments}
\end{deluxetable*}
\end{samepage}


\section{Velocity dispersion Computation}\label{sec:veldisp}

\subsection{Star Cluster center}\label{sec:cc}

The position of the cluster center affects the accuracy with which the velocity dispersion profile describes the dynamics of the cluster. In this work, we determine the cluster center in the 2D space by identifying the point of maximum density, calculated iteratively using the \texttt{Meanshift} algorithm \citep{meanshift}. We obtain the median position of all cluster members as the initial center and apply a first-pass filtering by circularly segmenting the 2D space using the half-mass radius, excluding stars located beyond this radius.
Each iteration moves the 2D cluster center towards regions of higher stellar density until convergence, and the coordinates of the center are computed as
\begin{equation}
    (\alpha_{\rm new}, \delta_{\rm new}) = 
    \frac{\sum_{i=1}^{n} (\alpha_i, \delta_i) \cdot f_X(\theta_i)}{\sum_{i=1}^{n} f_X(\theta_i)} \label{eq:iterat}
\end{equation}
where $n$ is the number of stars within the half-mass radius of the cluster; \(f_X(\theta_i)\) is the weight of members; (\(\alpha_i, \delta_i\)) is member 2D position in right ascension and declination; and (\(\alpha_{\rm new}, \delta_{\rm new}\)) is the new center position. We estimate the 3D center based on the 2D center using the coordinate transformation method. The 2D and 3D center of each cluster are listed in columns 3--7 in Table~\ref{tab:fdim}. 
The weight of each member in Eq.~\ref{eq:iterat} is computed using a Gaussian probability density function,
\begin{equation}
    f_X(\theta_i) = \frac{1}{\sqrt{2 \pi \sigma^2}} 
    \exp \left( -\frac{\theta_i^2}{2 \sigma^2} \right)
    \label{eq:gauss}
\end{equation}
where \(\theta_i\) is the angular distance of the star from the initial center, \(\sigma^2\) is the bandwidth.
Within the \texttt{Meanshift} framework, the bandwidth parameter defines the width of the Gaussian kernel and directly influences the resolution of the density estimate. To determine the optimal bandwidth, we employ cross-validation based on the least mean square error (LMSE) criterion. The bandwidth yielding the minimum mean square error (MSE) is selected, where the MSE quantifies the displacement between the initial position and the drifted position across iterations.


\subsection{One-Dimensional Velocity Dispersion}\label{sec:Vdisp}

We compute the dispersion of RVs and PMs of each cluster, to quantify their dynamical states. The likelihood function of the PM distribution can be modeled with two Gaussian functions \citep{pang2018}: cluster members \(L_{c}\) and field stars \(L_{f}\).
\begin{equation}
    L_{\rm tot} = n_{c} \cdot L_{c,i} + (1 - n_{c}) \cdot L_{f,i}
    \quad ,
\end{equation}
where
\begin{equation}
    L_{c,i} = \frac{1}{\sqrt{2\pi(\sigma_i^2 + \sigma_c^2)}} \exp\left(-\frac{(v_i - \mu_c)^2}{2(\sigma_i^2 + \sigma_c^2)}\right)
\end{equation}
and
\begin{equation}
    L_{f,i} = \frac{1}{\sqrt{2\pi(\sigma_i^2 + \sigma_f^2)}} \exp\left(-\frac{(v_i - \mu_f)^2}{2(\sigma_i^2 + \sigma_f^2)}\right)
    \quad , 
\end{equation}
where \(L_{\rm tot}\) is the total likelihood function; \(n_{c}\) is the fraction of the cluster component, equal to 95\%; \(v_i\) is the observed velocities for single stars and \(\sigma_i\) is the measurement uncertainties; 
\(\mu_c\) and \(\mu_f\) are the mean velocity of cluster members and field stars; and \(\sigma_c\) and \(\sigma_f\) are the velocity dispersions of cluster members and field stars. 
The optimal value of PM dispersion and the associated uncertainty are obtained by the Markov Chain Monte Carlo (MCMC) method (columns 8-9 in Table~\ref{tab:fdim}). 
During the fitting process, we set the initial values of \(\mu_c\) and \(\mu_f\) equal to the observed average $\mu_{\alpha}\cos\delta$ (or $\mu_{\delta}$) of the cluster. Similarly, the initial values of \(\sigma_c\) and \(\sigma_f\) are set to the observed standard deviations of $\mu_{\alpha}\cos\delta$ (or $\mu_{\delta}$).

The RV distribution follows a very similar likelihood function as the PMs distribution, incorporating the effects of binary broadening (equations~1 and~8 in \citealt{cottaar2012}). The Gaussian distribution of cluster members is broadened by the orbital motions of unresolved binary systems, and also by the uncertainties in the RV measurement \citep{pang2021a}. We simulate the broadening effects caused by binary stars. In the simulations, we adopt a log-normal orbital period distribution for the binaries \citep{raghavan_survey_2010}. For simplicity, we assume that the mass ratio of the binaries is uniformly distributed between 0 and 1 \citep{duchene_stellar_2013}, and that the eccentricity of the binary system is uniformly distributed between 0 and the maximum value \citep{parker2009binary}. Although this choice of parameters represents a simplified model, it provides a reasonable approximation of the characteristics of actual binary stars. Table~\ref{tab:fdim} lists the final RV dispersion (column 10). 


\subsection{One-Dimensional Velocity Dispersion Profile}\label{subsec:1Dpro}

In order to derive velocity dispersion profiles from cluster member's PMs, we divide the clusters into subgroups with the same number of members based on the distance from each cluster member to the cluster center. We specify that each subgroup has more than 20 members, and the total number of subgroups should be between 10 and 20. In cases where equal division is not feasible, we choose the scheme that has the largest ratio between the smallest and largest subgroup sizes. And the total number of subgroups is closest to 15. 
Subsequently, the PM dispersion for each subgroup is calculated using the same method outlined in Section~\ref{sec:Vdisp}. The distance from the subgroup to the cluster center is the average distance of each member. 

We adapted the method for obtaining the RV dispersion profile from that of the PMs dispersion profile, due to the limited RV coverage in Gaia DR\,3.
Only those with accurate RV measurements are selected to participate in the subgroups. To attenuate the influence of binaries on the RV dispersion profile, we further screen the members centered around the average RV  within a range of three times the median absolute deviation (MAD). Due to the small number of members with accurate RV measurements, the acceptable total number of subgroups ranges from 4 to 20, with each subgroup containing more than 15 members.

\begin{figure}[tb!]
\centering
\includegraphics[angle=0, width=1\textwidth]{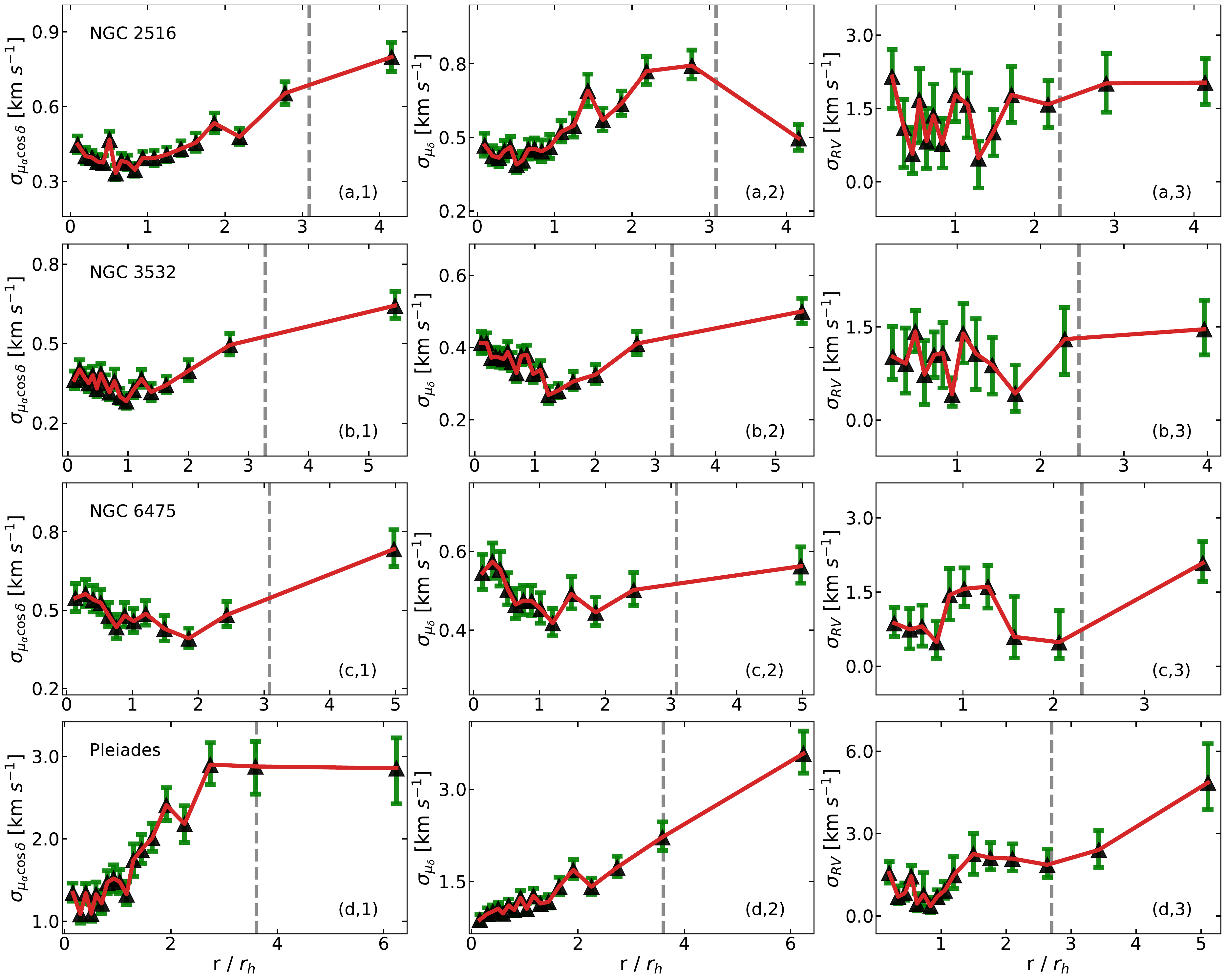}
    \caption{The $\mu_{\alpha}\cos\delta$ ($\mu_{\delta}$) and RV dispersion profiles of the four clusters: NGC\,2516, NGC\,3532,  NGC\,6475 and Pleiades. The red solid curve represents the 1D velocity dispersion profile. The black triangles represent the velocity dispersion of each subgroup. Distances from each subgroup to the cluster center are normalised using the half-mass radius \(r_h\). The green bars are the errors in velocity dispersion for each subgroup. The grey dashed line represents the tidal radius of each cluster.}
\label{fig:vdisp_prof_1}
\end{figure}

\begin{figure}[tb!]
\centering
\includegraphics[angle=0, width=1\textwidth]{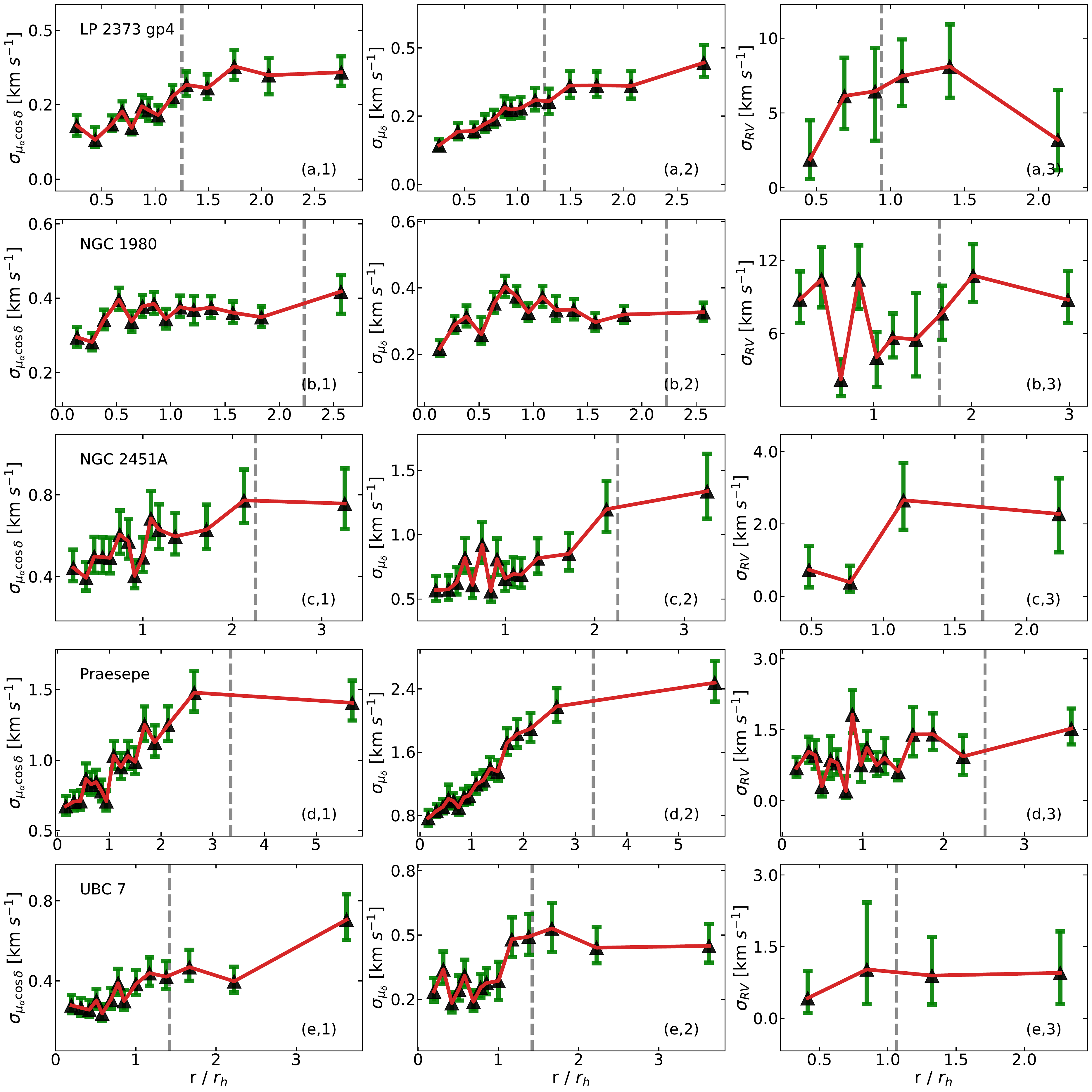}
    \caption{The $\mu_{\alpha}\cos\delta$ ($\mu_{\delta}$) and RV dispersion profiles of the five clusters: LP\,2373\,gp4, NGC\,1980, NGC\,2451A, Praesepe and UBC\,7. Colors and symbols are the same as in Fig.~\ref{fig:vdisp_prof_1}.}
\label{fig:vdisp_prof_2}
\end{figure}

The 1D dispersion profiles of the nine clusters (PM components and RV) are shown in Figs.~\ref{fig:vdisp_prof_1} and~\ref{fig:vdisp_prof_2}. By comparing the three velocity dispersion profiles of each cluster, we find that the velocity dispersion of the outermost region (outside the tidal radius, grey dashed vertical dashed line) of NGC\,2516 in the $\mu_{\delta}$ and LP\,2373gp4 in the RV is significantly lower than its inner parts (panel (a,2) in Fig.~\ref{fig:vdisp_prof_1} and panel (a,3) in Fig.~\ref{fig:vdisp_prof_2}). However, this phenomenon does not occur in the other two directions. This discrepancy may be attributed to uneven tidal forces acting on the cluster. Limited by Gaia RV precision, the RV dispersion profiles have larger error bars. On average 26\% of members in our target clusters have Gaia RV observations. Only Pleiades and NGC\,6475 show an increasing trend of $\sigma_{\rm RV}$ toward larger radius. The other six clusters do not exhibit a clear or consistent overall trend in their RV dispersion profiles.

The dispersion profiles of $\mu_{\alpha}\cos\delta$ and $\mu_{\delta}$ for LP\,2373\,gp4, NGC\,2451A, Pleiades, Praesepe and UBC\,7 exhibit a pronounced increasing trend with radial distance. This trend may indicate ongoing tidal disintegration under the influence of the Galactic tidal field, while stars are escaping from these clusters. The $\mu_{\alpha}\cos\delta$ ($\mu_{\delta}$) dispersion profiles for NGC\,1980 are relatively flat, with only small differences in velocity dispersion between adjacent subgroups. This trend may indicate that NGC\,1980 is approaching dynamical equilibrium. 
The $\mu_{\alpha}\cos\delta$ ($\mu_{\delta}$) dispersion profiles for NGC\,2516 (except $\mu_{\delta}$), NGC\,3532 and NGC\,6475 reach a maximum near the cluster center, decreasing within 1\,$r_h$, then increase towards larger radii. This trend may be triggered by strong mass segregation under clumpy and sub-virial initial conditions \citep{mario2016}, or by the presence of stellar-mass BHs in the central regions of these clusters (\citealt{torniamenti_stellar-mass_2023}; \citealt{Aros2023}; \citealt{Gottgens2021}), which could both enhance velocity dispersion in the central region. 

\begin{figure}[tb!]
\centering
\includegraphics[angle=0, width=1\textwidth]{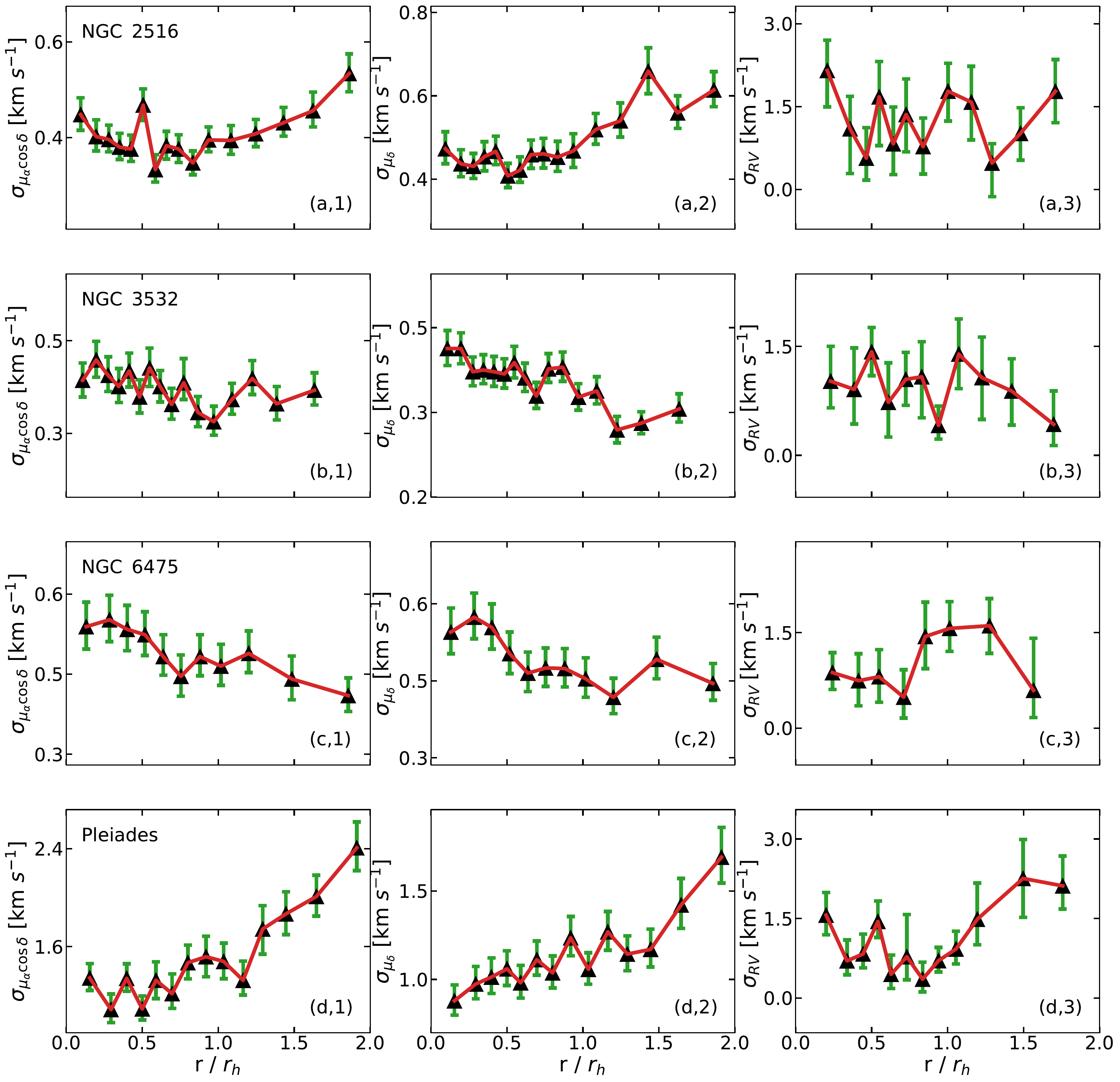}
    \caption{The $\mu_{\alpha}\cos\delta$ ($\mu_{\delta}$) and RV dispersion profiles with $r<2\,r_h$ of the four clusters: NGC\,2516, NGC\,3532,  NGC\,6475 and Pleiades. Colors and symbols are the same as in Fig.~\ref{fig:vdisp_prof_1}.}
\label{fig:vdisp_prof_in}
\end{figure}

In Fig.~\ref{fig:vdisp_prof_in} we show the dispersion profiles of $\mu_{\alpha}\cos\delta$, $\mu_{\delta}$ and the RV for NGC\,2516, NGC\,3532 and NGC\,6475 and Pleiades, within $2\,r_h$, which is inside the tidal radius of all these clusters. 
NGC\,6475 shows the clearest trend: a central peak within 1\,$r_h$ in the velocity dispersion profile in $\mu_{\alpha}\cos\delta$ (Fig.~\ref{fig:vdisp_prof_in}\,(a)). The $\sigma_{\mu_{\alpha}\cos\delta}$ and $\sigma_{\mu_{\delta}}$ profiles of NGC\,2516 and NGC\,3532 increase towards the center ($<0.5\,r_h$), and then increase toward larger radii. This implies that in these two clusters the processes of two-body relaxation and tidal stripping are interacting, and influence each other. 
On the other hand, Pleiades shows a consistent increasing trend in both $\sigma_{\mu_{\alpha}\cos\delta}$ and $\sigma_{\mu_{\delta}}$. This indicates ongoing disruption, which is consistent with the fact that Pleiades has already developed tidal tails (\citealt{Bhattacharya2022}; \citealt{pang2022a}). Two clusters, LP\,2373\,gp4 and NGC\,1980, have significantly larger RV dispersions due to poorer measurement quality. The RV uncertainties in these two clusters are the highest, averaging up to 6.4~$\rm km\,s^{-1}$, which is several times larger than those of other clusters.

\subsection{Dependence of the Velocity Dispersion on Stellar Mass}

As the cluster evolves, its overall velocity dispersion gradually decreases, due to cluster expansion and mass loss. High-mass stars transfer kinetic energy to low-mass stars through two-body relaxation; therefore, the  velocity dispersion of the high-mass stars is reduced. During this process, low-mass stars gain kinetic energy and migrate to the outskirts of the cluster, exhibiting higher velocity dispersion \citep{Aros2023}. In the full energy equipartition state or during an expansion, we expect to observe a decreasing trend of velocity dispersion with increasing stellar mass \citep{Trenti2013}.

\begin{figure}[tb!]
\centering
\includegraphics[angle=0, width=1\textwidth]{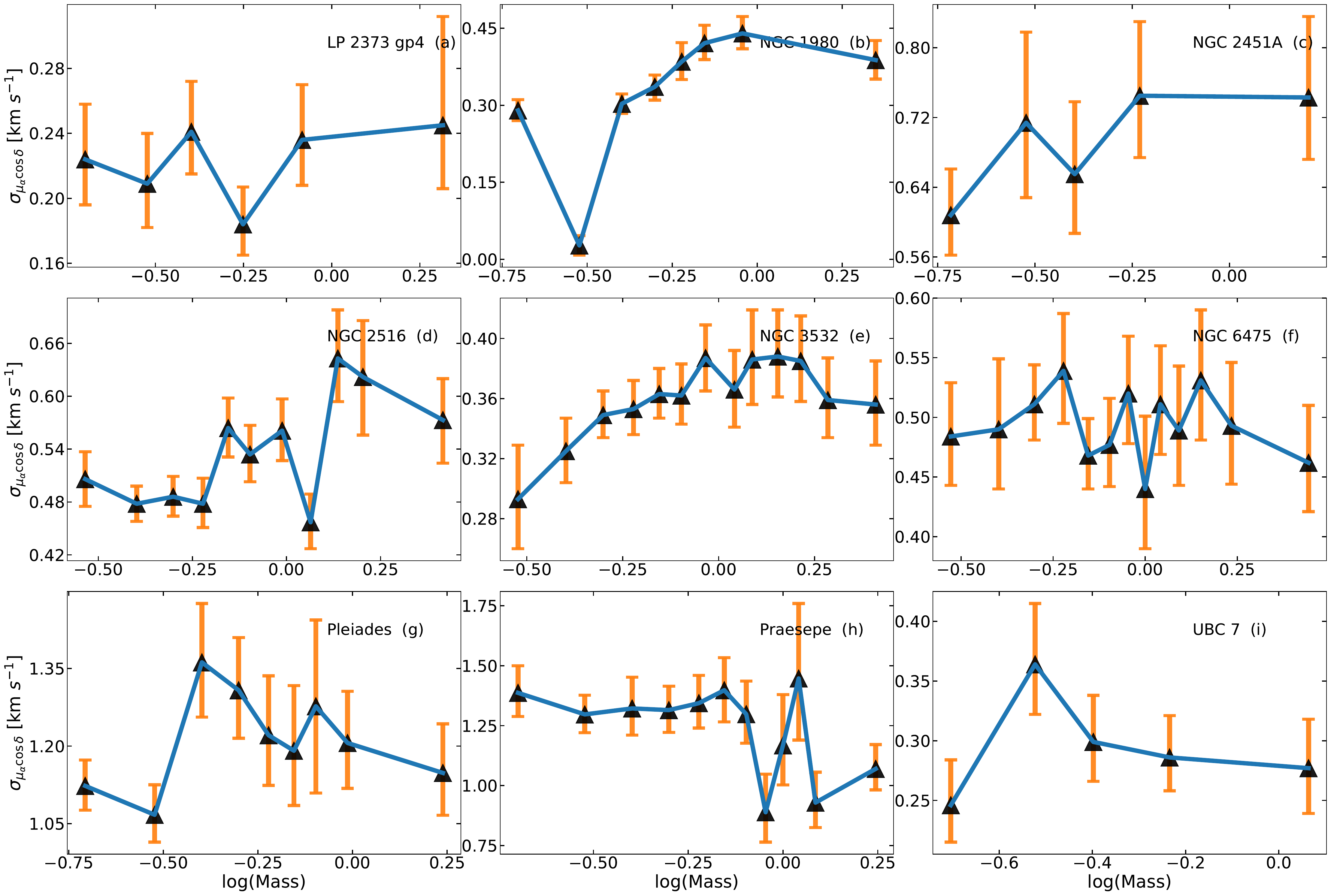}
    \caption{The $\mu_{\alpha}\cos\delta$ dispersion profiles of the nine clusters along stellar mass mass {\bf within tidal radius}: LP2373\,gp4, NGC\,1980, NGC\,2451A, NGC\,2516, NGC\,3532, NGC\,6475, UBC\,7, Praesepe, and Pleiades. The blue solid curve represents the 1D velocity dispersion profile. The black triangles represent the velocity dispersion of each subgroup. Masses from each subgroup are expressed in logarithmic scale. The orange bars represent the errors in the velocity dispersion for each subgroup.}
\label{fig:mass_sigma_ra}
\end{figure}

\begin{figure}[tb!]
\centering
\includegraphics[angle=0, width=1\textwidth]{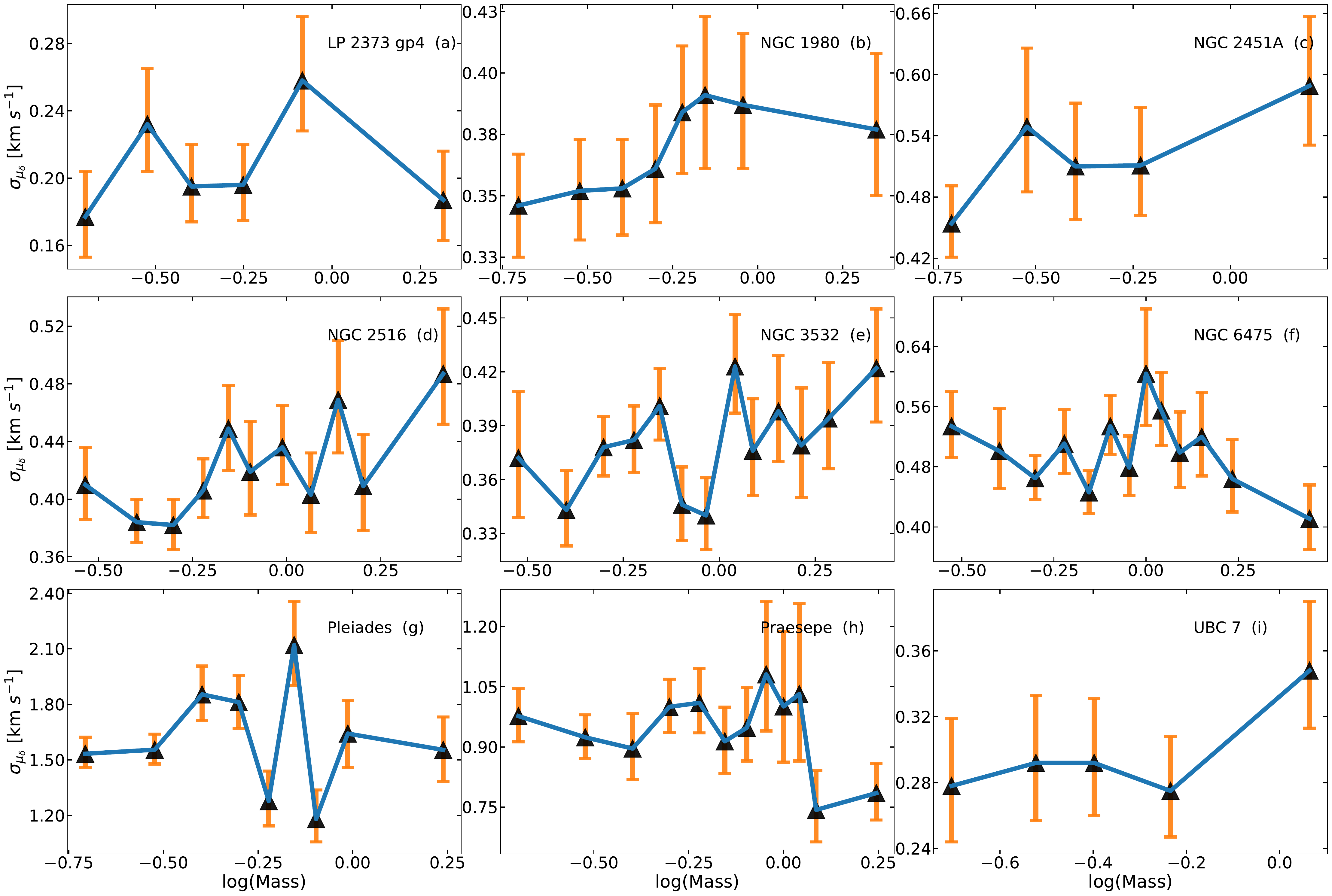}
    \caption{The $\mu_{\delta}$ dispersion profiles of the nine clusters along stellar mass mass within tidal radius: LP2373\,gp4, NGC\,1980, NGC\,2451A, NGC\,2516, NGC\,3532, NGC\,6475, UBC\,7, Praesepe, and Pleiades. Colors and symbols are the same as in Fig.~\ref{fig:mass_sigma_ra}.}
\label{fig:mass_sigma_dec}
\end{figure}

\begin{figure}[tb!]
\centering
\includegraphics[angle=0, width=1\textwidth]{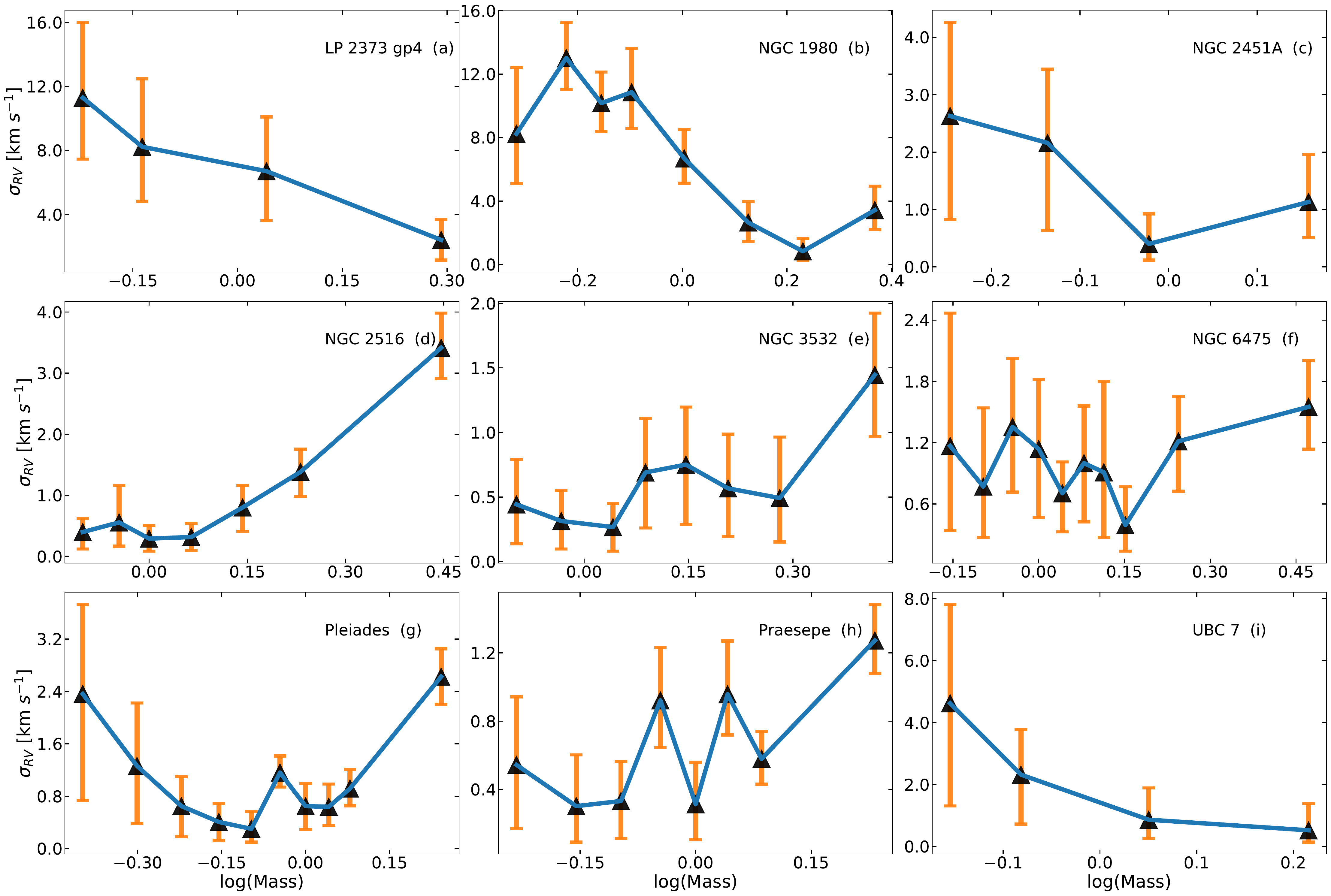}
    \caption{The RV dispersion profiles of the nine clusters along stellar mass within tidal radius: LP2373\,gp4, NGC\,1980, NGC\,2451A, NGC\,2516, NGC\,3532, NGC\,6475, UBC\,7, Praesepe, and Pleiades. Colors and symbols are the same as in Fig.~\ref{fig:mass_sigma_ra}.}
\label{fig:mass_sigma_rv}
\end{figure}

We display the dependence of velocity dispersions $\sigma_{\mu_{\alpha}\cos\delta}$, $\sigma_{\mu_{\delta}}$, and $\sigma_{\rm Rv}$ on stellar mass for stars within the tidal radius of each cluster in Figures~\ref{fig:mass_sigma_ra}, \ref{fig:mass_sigma_dec}, and~\ref{fig:mass_sigma_rv} respectively. 
A declining dependence of velocity dispersion on stellar mass is observed in several clusters. NGC\,6475 and Praesepe both show decreasing values of $\sigma_{\mu_{\alpha}\cos\delta}$ and $\sigma_{\mu_{\delta}}$ with increasing stellar mass. In LP\,2373\,gp4, $\sigma_{\mu_{\delta}}$ and $\sigma_{\rm Rv}$ also decrease when the stellar mass increases. Hence, these three clusters might be undergoing energy equipartition or expansion. 

In NGC\,2516 and NGC\,3532, on the other hand, $\sigma_{\mu_{\alpha}\cos\delta}$, $\sigma_{\mu_{\delta}}$ and $\sigma_{\rm Rv}$ all increase with increasing stellar mass. Some dynamical processes must have occurred in these two clusters, preferentially ejecting massive stars. This behavior is contrary to what is expected from energy equipartition. When massive stars sink toward the center of the cluster via two-body relaxation, they are expected to obtain a lower velocity dispersion. 
$N$-body simulations by \citet{mario2016} show that clumpy sub-virial initial conditions can lead to strong mass segregation processes. Under these conditions, the most massive stars or stellar remnants have higher velocity dispersion than the lower-mass stars, which might attribute to the observed profile in NGC\,2516 and NGC\,3532. 
 However, \citet{pang2023} found no clear evidence of mass segregation in NGC\,2516 and NGC\,3532. This suggests that strong mass segregation triggered by clumpy sub-virial initial conditions cannot account for the observed higher velocity dispersion in high-mass stars in these two clusters. The age of NGC\,2516 (123\,Myr) is about one third of its relaxation timescale (323\,Myr), while the age of NGC\,3532 (398\,Myr) is slightly older than its relaxation time (265\,Myr), computed using Equation (7.108) in \citet{binney2008} adopting the
half-mass radius, the total mass (Table~\ref{tab:fdim}), and the number of members. Considering the uncertainty in age estimation, NGC\,2516 and NGC\,3532 are not relaxed yet.
Another possible cause of the observed higher velocity dispersion of high-mass stars in NGC\,2516 and NGC\,3532 could be due to stellar interactions involving binaries or BHs in the center that expel massive stars, resulting in a higher velocity dispersion. 

We obtain an approximation of the virial state of each cluster by estimating the ratio of the dynamical mass \citep{fleck2006} over the photometric mass. When the ratio is significantly larger than 1 \citep{pang2021a}, the cluster is supervirial. Most of the clusters are supervirial, having a ratio value between 5 and 70. Considering the incompleteness of the photometric mass of observed clusters below 0.25 solar mass (Gaia observation limit), the ratio is still larger than 1. However, there are two exceptions: NGC\,2516 and NGC\,3532. These two clusters have the lowest virial ratios, 4 and 2 respectively. Given a missing mass of 10\% from low-mass stars due to incompleteness from Gaia observations \citep{tang2019}, we conclude that NGC\,2516 and NGC\,3532 have just started to depart from virial equilibrium, consistent with their not-relaxed dynamical state.

\section{$N$-body simulations of model open clusters}\label{sec:nbody}

To interpret our observational results, we make a comparison with OC models using the direct $N$-body approach \citep[e.g.,][]{Aarseth_2003}. The initial density and kinematic profiles of the clusters are according to the \citet{Plummer_1911} model. We model star clusters with initial masses of $\mcl(0)=2\,000\,\Ms$, $5\,000\,\Ms$ and $10\,000\,\Ms$, and initial half-mass radii of $\rh(0)=1.0$~pc and 2.0~pc. All models assume solar metallicity and are subjected to a solar-neighbourhood-like external Galactic field.

The initial models comprise zero-age main-sequence (ZAMS) stars with masses $0.08\,\Ms\leq m_\ast\leq150.0\,\Ms$, that are distributed according to the canonical initial mass function \citep[IMF;][]{kroupa2001}.
The overall (initial) primordial-binary fraction is taken to be $\fbin(0)=30$\%, which is typical for open clusters. However, the binary fraction of O-type stars ($m_\ast\geq16.0\,\Ms$), which are initially paired only among themselves, is $\fobin(0)=100$\%, consistent with the observed high binary fraction among O-stars in young clusters and associations \citep[e.g.][]{Sana_2011,Moe_2017}. The O-star binaries initially follow the observed orbital-period and eccentricity distributions of \citet{Sana_2011} and a uniform mass-ratio distribution. The lower mass binaries initially follow the orbital period distribution of \citet{Duq_1991}, a uniform mass ratio distribution, and the thermal eccentricity distribution.

In order to compare with observations, all model clusters are evolved for $\approx 500$\,Myr. This timescale ensures that the models remain within the typical age range of the majority of observed clusters. Considering the oldest cluster in the sample ($\approx 700$\,Myr), we extend the evolution of the $\mcl(0)=2\times10^3\,\Ms$, $\rh(0)=2$~pc models to $\approx 800$~Myr.

For comparison, we also evolve several $\mcl(0)=5\times10^3\,\Ms$, $\rh(0)=2$~pc models without primordial binaries. To reduce statistical uncertainty in the computed models, we perform multiple simulations for each set of initial parameters, with the initial models realized using different random seeds. The specific parameters for each model can be found in Table~\ref{tab:nmodel}. 

The model clusters are evolved using the direct $N$-body code {\tt NBODY7} \citep{Aarseth2012}. {\tt NBODY7} is a variant of {\tt NBODY6} \citep{Nitadori_2012} that utilizes graphical processing units (GPU) to accelerate the regular force calculations and parallel CPU threads to evaluate the irregular forces. The singularity due to the diverging gravitational force between the members of a hard binary is dealt with by applying the Kustaanheimo-Stiefel regularization (KS-regularization). We refer to \cite{Aarseth_2003,Aarseth2012} for details of the various algorithms used in {\tt NBODY6} and {\tt NBODY7}. The main difference between {\tt NBODY6} and {\tt NBODY7} is that the latter applies the Algorithmic Regularisation Chain ({\tt ARCHAIN}) \citep{Mikkola_1999} instead of the classical {\tt KS-Chain} \citep{Mikkola_1993}, to regularize triple and higher-order hierarchical systems. For compact hierarchical systems comprising high mass ratios, that can frequently form via dynamical interactions in high-binary-fraction clusters like those we model, {\tt ARCHAIN} has proven to be generally more accurate and stable. 

{\tt NBODY7} incorporates stellar and binary evolution by coupling to the semi-analytical binary evolution code {\tt BSE} \citep{Hurley_2002}. This allows us to take into account the effects of mass loss and remnant formation for every stellar and binary-star member of the cluster. In our computations, we adopt the \citet{Vink_2001} stellar wind model, BH and neutron star (NS) formation via the `rapid' remnant mass and fallback models of \citet{Fryer_2012}, and the momentum-conserving remnant natal kick. We refer to \cite{banerjee2020,Banerjee_2020c} for detailed descriptions and applications of these updates to {\tt NBODY7}. Owing to the low escape speeds of the presently modeled clusters, only a few BHs and NSs are retained in the clusters, and they do not result in any general-relativistic mergers.

In this study, our goal is not to replicate any specific open cluster. Instead, we compare the observed trends with those derived from simulations to identify the underlying physical processes (e.g., two-body relaxation, tidal fields) that may drive the observed patterns. Accurately reproducing a single cluster lies beyond the scope of this paper.


\section{Comparison between observations and simulations}\label{sec:comparison}

In order to accurately identify binary systems in the output generated by the $N$-body simulations, we follow the approach outlined below. For each pair of mutually nearest neighbouring stars, we calculate the semi-major axis, \(a_{ij}\), of the two-body system:
%
%
%
\begin{equation}
    a_{ij} = \left(
        \frac{2}{r_{ij}} - \frac{v_{ij}^2}{G(m_i+m_j)}
        \right)^{-1}
    \label{eq:sma}
\end{equation}
where \(r_{ij}\) is the distance between the two stars, and \(v_{ij}\) is the relative velocity; \(G\) is the gravitational constant, and \(m_i+m_j\) is the sum of the masses of the two stars. Pairs with a positive semi-major axis (i.e., with a negative binding energy) are identified as a binary system, while others are considered single stars. 

Binary systems often remain unresolved due to the spatial resolution of Gaia observations \citep{Castro2024}. Unresolved binaries in observational surveys are typically treated as single stars. The angular resolution of 0.6 arcsecond from Gaia DR\,3 corresponds to a semi-major axis of 300\,AU for a binary system at a distance of 500\,pc \citep[e.g.,][]{pang2023}. 
To ensure consistency between simulations and observations from Gaia DR\,3, we treat identified binaries in the model cluster whose semi-major axis less than 300\,AU  as unresolved systems. The velocity and spatial position of each identified binary are represented as the values at the center of mass of the two-body system.

In the Model~2 (see Table~\ref{tab:nmodel}), we identify two stars with high radial velocities: one with ${\rm RV} = 128.3\,{\rm km\,s}^{-1}$ and the other with ${\rm RV} = -44.5\,{\rm km\,s}^{-1}$. These two stars remain in close proximity at all times, and the pair has a positive semi-major axis, indicating that this could be a transient bound binary system. Such high velocities could be produced by a past three-body interaction, where one star is ejected while the remaining two form a transient binary. To eliminate potential multiple systems, we filter out stars with velocities that lie beyond $\pm 3\sigma$ from the mean velocity within each subgroup, where the subgroups are defined by cluster-centric distance.

To compare the 1D velocity dispersion profiles of the models with those of the observed clusters, we first exclude stars in the model clusters with masses below the observed cluster's mass limit of Gaia DR\,3 (column~13 in Table~\ref{tab:fdim}). Next, we apply the same plotting method used for the observed clusters (see Section~\ref{subsec:1Dpro}), fixing the number of subgroups to 20. The 1D PM dispersion for each subgroup is then calculated using Eq.~\ref{eq:1DPM}. Finally, the 1D velocity dispersion profile for each model cluster is obtained by averaging the simulation results of different random seeds for that model.
\begin{equation}
    \sigma_{\rm 1DPM}(r) = \frac{1}{2} \sqrt{\sigma_{\mu_{\alpha}\cos\delta}^2(r) + \sigma_{\mu_{\delta}}^2(r)}
    \quad ,
    \label{eq:1DPM}
\end{equation}
where $r$ is the projected distance between the center and subgroup.

\begin{figure}[tb!]
\centering
\includegraphics[angle=0, width=1.00\textwidth]{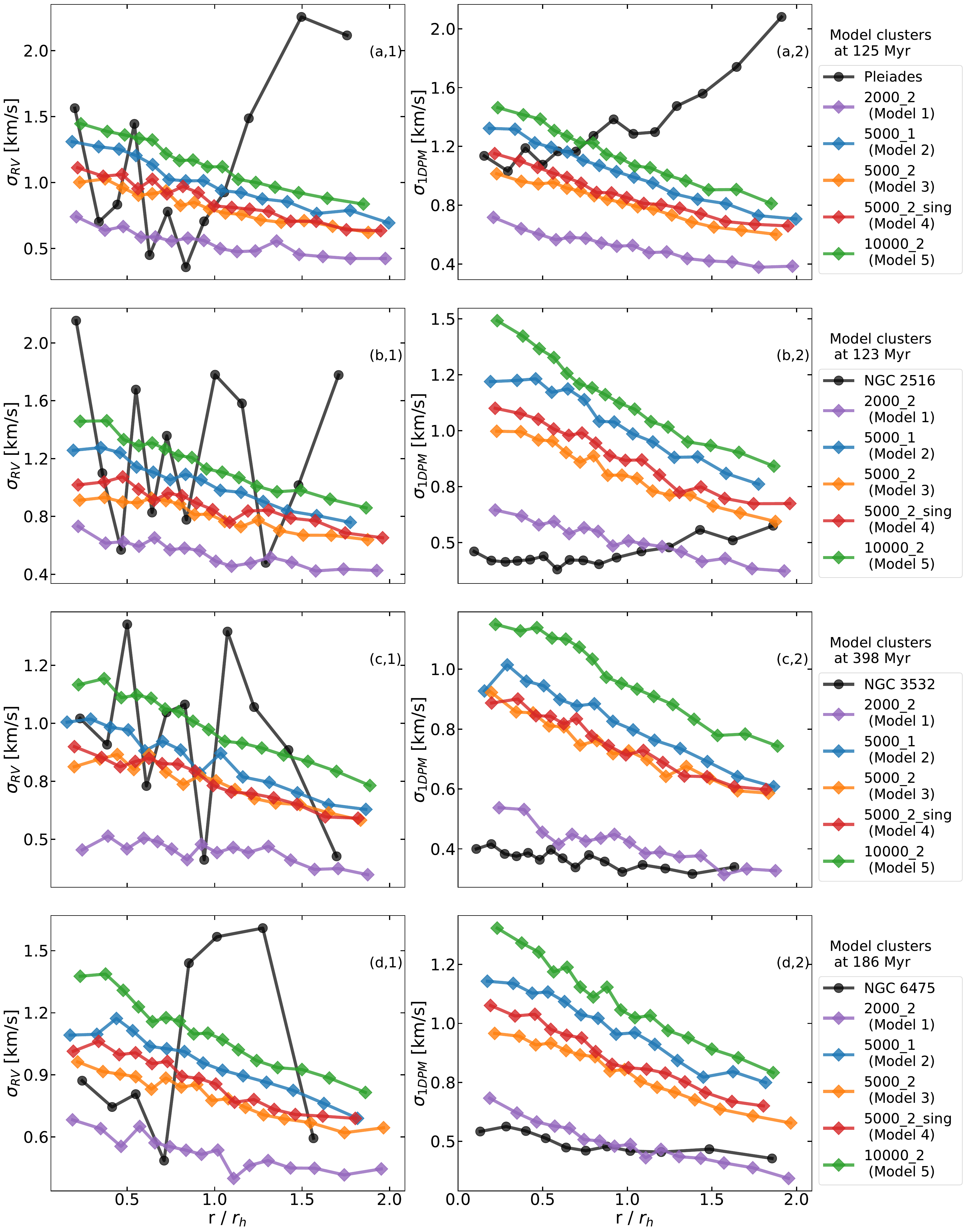}
    \caption{The PM and RV dispersion profiles of four observed clusters and model clusters at specific evolutionary stages are presented. The black solid curves represent the 1D velocity dispersion profiles of the observed clusters, while the colored solid curves correspond to the 1D velocity dispersion profiles of the different model clusters. Distances from each subgroup to the cluster center are normalized using the half-mass radius \(r_h\). Only the 1D velocity dispersion profiles for $r \leq 2r_h$ are shown.}
\label{fig:nbody0}
\end{figure}

We compare the 1D velocity dispersion profiles of all model clusters with those of each observed cluster over evolutionary durations from 0~Myr to 500~Myr in steps of 50~Myr. 
Fig.~\ref{fig:nbody0} presents the fitting results for the model cluster at an age consistent with that of the observed cluster.
Both RV and PM dispersion profiles exhibit a rising trend towards the cluster center. Model clusters with different initial conditions display similar profile shapes at the same evolutionary duration. 
In all panels of Fig.~\ref{fig:nbody0}, the Model~2 (see Table~\ref{tab:nmodel}) shows higher dispersion values than the Model~3 (see Table~\ref{tab:nmodel}). Both have the same mass but Model~3 has a larger half-mass radius. This suggests that among clusters with identical initial mass, those with greater central density exhibit  a higher velocity dispersion across their profiles, consistent with expectations for systems near virial equilibrium \citep[][]{spitzer2014, heggie2003}.

A comparison between the RV dispersion profiles of the four observed clusters and those of the model clusters reveals a clear difference. The observed clusters do not follow a consistent trend and display a wider range of dispersion values. This discrepancy may result from incomplete RV datasets, making it difficult to construct accurate one-dimensional RV dispersion profiles.

\begin{figure}[tb!]
\centering
\includegraphics[angle=0, width=1.00\textwidth]{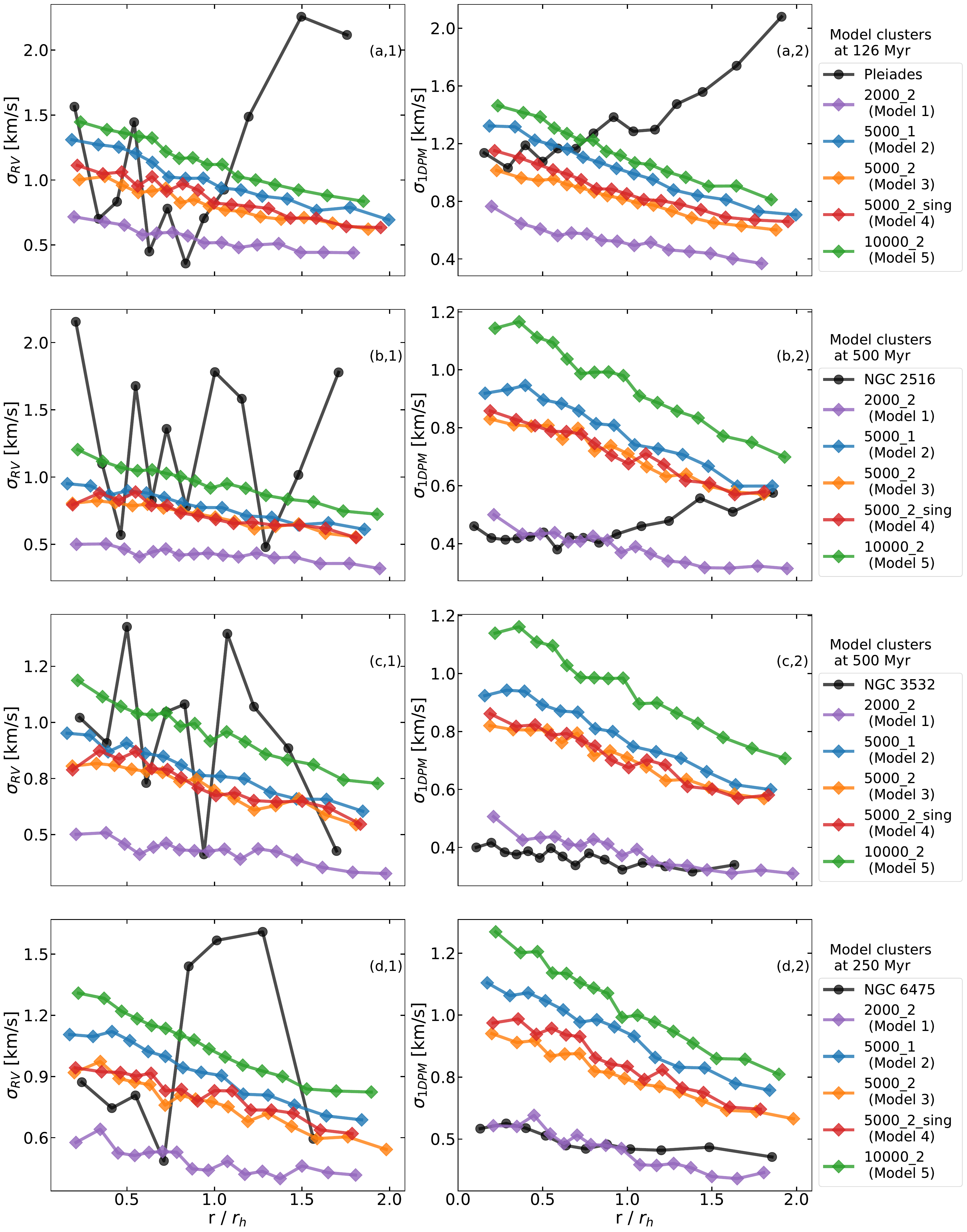}
    \caption{The PM and RV dispersion profiles of four observed clusters and model clusters at specific evolutionary stages are presented. Colors and symbols are the same as in Fig.~\ref{fig:nbody0}.
    }
\label{fig:nbody1}
\end{figure}

For the 1D PM dispersion profile in Fig.~\ref{fig:nbody1}, Pleiades exhibits an overall increasing trend away from the cluster center (panel a2), which is inconsistent with the profiles of the model clusters. We discuss the possible reasons for this result in detail in Section~\ref{sec:dis}. 
For NGC\,2516, NGC\,3532 and NGC\,6475, their 1D PM dispersion profiles clearly lie below those of model clusters evolved to their respective ages (panels~b2, c2, and d2). The fitting of the model clusters using the same evolutionary timescales as the observations is relatively poor.

To identify the model clusters that best fit the 1D velocity dispersion profile of the observed clusters, along with their corresponding evolutionary times, Fig.~\ref{fig:nbody1} presents the best-fit results from the model clusters with varying initial conditions. The best-fit results are obtained based on eye-fitting when the model cluster whose 1D velocity dispersion profile closely matches that of the observed cluster.

The profile of NGC\,6475 fits best with the Model~1 (see Table~\ref{tab:nmodel}) evolved to 250~Myr (panel~d2), a significant overlap in the range of distances less than \(0.75 r_h\).  
The 1D PM dispersion profiles of NGC\,2516 and NGC\,3532 are consistently lower than those of individual model clusters up to the age 
500~Myr (panels~b2 and~c2). To enable meaningful comparisons, we therefore require model clusters with longer evolutionary durations. Among the available models, only Model~1 (see Table~\ref{tab:nmodel}) has been evolved beyond 500~Myr, we will focus our further studies on these model cluster data.

\begin{figure}[tb!]
\centering
\includegraphics[angle=0, width=1.05\textwidth]{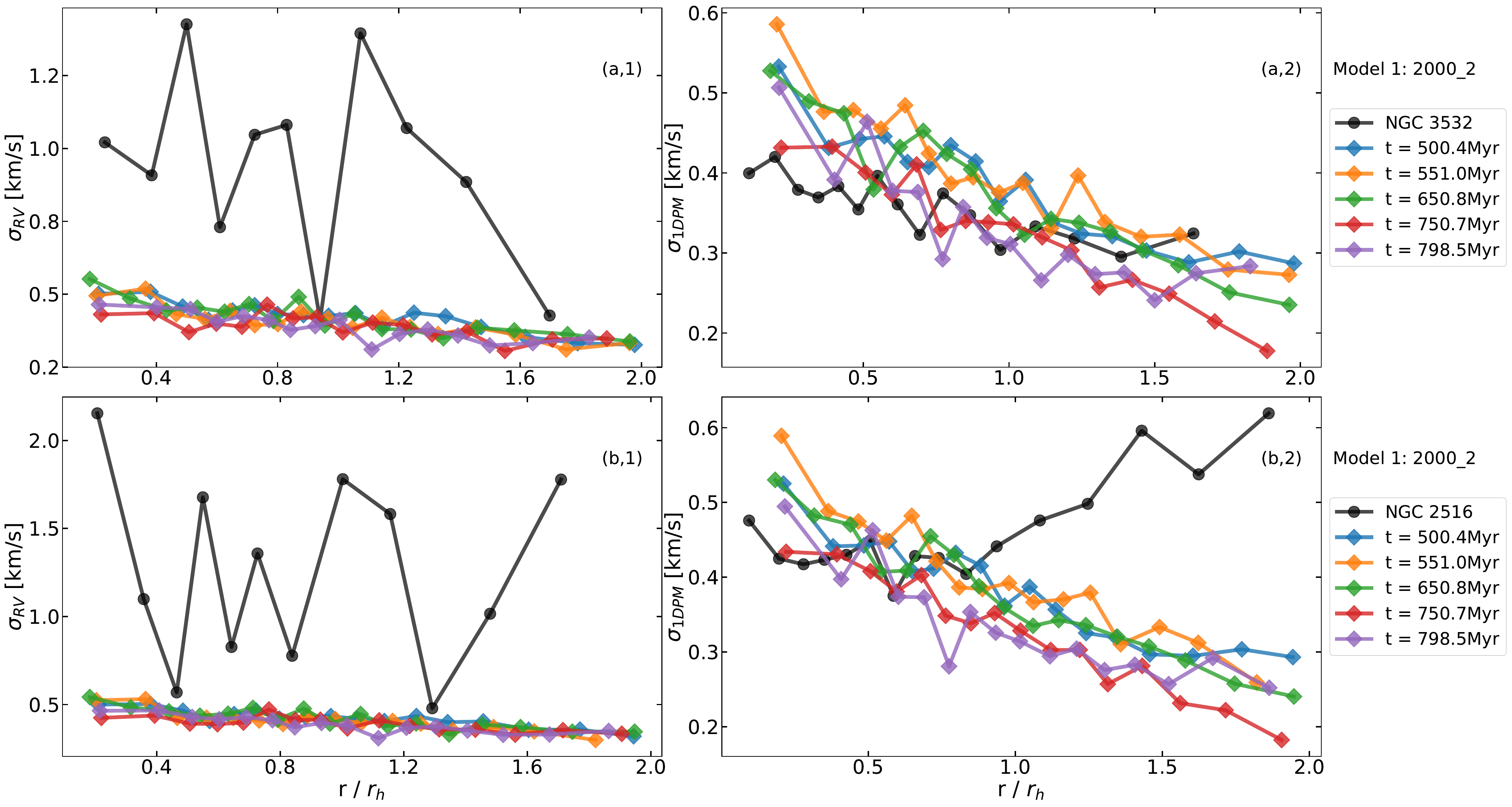}
    \caption{The PM and RV dispersion profiles of NGC\,3532 and NGC\,2516. The black solid curves represent the 1D velocity dispersion profiles of the observed clusters, while the colored solid curves correspond to the 1D velocity dispersion profiles of Model 1 ($\mcl(0)=2\times10^3~\Ms$, $\rh(0)=2$~pc)  at different evolutionary times.}
\label{fig:nbody2}
\end{figure}

Fig.~\ref{fig:nbody2} illustrates the 1D PM dispersion profiles of NGC\,3532 and NGC\,2516 compared to those of the Model~1 (see Table~\ref{tab:nmodel}), which evolved from 500~Myr to 800~Myr. In Fig.~\ref{fig:nbody2}a, the Model~1 (see Table~\ref{tab:nmodel}) exhibits a clear decrease in the overall velocity dispersion as the evolutionary duration increases. The dispersion profile at 500~Myr lies entirely above that at 750~Myr, indicating that, under identical initial conditions, model clusters tend to have lower internal velocity dispersions at later evolutionary stages.
The decrease in the velocity dispersion with evolutionary time occurs in a cluster due to its expansion that is driven by dynamical relaxation and mass loss.

Also in Fig.\ref{fig:nbody2}a, the PM dispersion profile of NGC\,3532 remains consistently lower than that of the 800~Myr model within the region of distances less than \(0.75 r_h\), suggesting that 
the current model does not adequately reproduce the observed kinematics of this cluster.
In Fig.~\ref{fig:nbody2}b, the PM dispersion profile of NGC\,2516 shows a rising trend beyond \(0.6 r_h\), which contrasts with the declining trends seen in the model clusters. This inconsistency further suggests that the kinematics of the outskirts of NGC\,2516 have already been affected by the Galactic tidal field.

\begin{samepage}
\startlongtable
\begin{deluxetable*}{C LCLRCC}
\tablecaption{Initial conditions for 5 model clusters. \label{tab:nmodel}}
	\tabletypesize{\scriptsize}
	\tablehead{
            \colhead{Model}    & 
            \colhead{$\mcl(0)$}       &
		\colhead{$\rh(0)$}      & 
            \colhead{$Z$}        &
            \colhead{$f_{\rm bin}$}        &
            \colhead{$t$}        &                     
		\\
		\colhead{}              & 
		\colhead{($\Ms$)}         & 
		\colhead{(pc)}         &
		\colhead{}          &
  	    \colhead{}          &
		\colhead{(Myr)}         &    
		\\
            \cline{2-6}
	    \colhead{(1)}   & 
	    \colhead{(2)}   & \colhead{(3)} & 
	    \colhead{(4)}   & \colhead{(5)} &
	    \colhead{(6)}   & 
		}
	\startdata

\mathrm{1} & 2000 & 2.0 & 0.02 & 0.3 & 800 \\

\mathrm{2} & 5000 & 1.0 & 0.02 & 0.3 & 500 \\

\mathrm{3} & 5000 & 2.0 & 0.02 & 0.3 & 500 \\

\mathrm{4} & 5000 & 2.0 & 0.02 & 0 & 500 \\

\mathrm{5} & 10000 & 2.0 & 0.02 & 0.3 & 500 

    \enddata
	\begin{tablecomments}{
       Initial conditions for the star cluster simulations. Columns~2 and~3 list the initial total mass and half-mass radius of the model clusters, respectively. Columns~4 to~6 provide the metallicity, binary fraction and the simulation time.}
    \end{tablecomments}
\end{deluxetable*}
\end{samepage}

\section{Discussion}\label{sec:dis}

Our model clusters do not fully fit the profiles of the Pleiades, NGC\,2516, NGC\,3532, and NGC\,6475 at the current ages. Due to the uncertainty of RV measurement from Gaia DR\,3, the RV dispersion profiles of the observed clusters have a large scatter and do not exhibit coherent trends that are seen in the model clusters.

The 1D PM dispersion profile of the Pleiades displays a trend that contrasts with that of the model, potentially due to the influence of its external environment, such as the Galactic tidal field. Stars located in the marginal regions of the cluster are clearer to external perturbations, which can elevate their velocities. Some of these stars may have attained escape velocity, thereby contributing to the increase in velocity dispersion towards the cluster outskirts.

Comparisons between observations and simulations show that velocity dispersions of the model clusters are much higher than observed ones. This discrepancy may be attributed to a few possible causes. First, a higher initial mass is adopted in the model clusters compared to the observed clusters. Second, another plausible reason might be that the observed cluster has undergone significant expansion after gas expulsion. Our simulations do not incorporate several physical processes, such as early residual gas expulsion or the impact of tidal interactions with the parent molecular cloud. These mechanisms are capable of significantly reducing the kinetic energy of the system during its early dynamical evolution. This would result in a lower velocity dispersion in observations than in simulations. Third, an alternative scenario is that the observed clusters retained more stellar-mass BHs than the model clusters. BHs can rapidly expel massive stars and heat up the system, hence accelerating the dynamical evolution.

The 1D PM dispersion profiles of NGC\,2516, NGC\,3532, NGC\,6475 remain consistently lower than those in the model clusters (Figure~\ref{fig:nbody0} and \ref{fig:nbody1}). However, the observed total mass of NGC\,2516 and NGC\,3532 is (around) above 2000~\Ms, thus initial mass of model clusters cannot be the cause for the apparent discrepancy for these two clusters (Figure~\ref{fig:nbody2}). However, for the model clusters with masses of 5000~\Ms or above, the velocity dispersions are substantially higher than those of the observations. We increase the age of Model~1 (2000~\Ms) to 800~Myr. In this case, the model cluster fits better with NGC\,3532 and NGC\,2516, respectively, although the actual ages of the clusters are 398~Myr and 123~Myr. Either gas expulsion or stellar-mass BH retention may have accelerated the dynamical evolution in observed clusters. The difference of the observed age and the age of the model cluster may imply a larger fraction of BH retention. 

\begin{figure}[tb!]
\centering
\includegraphics[angle=0, width=1.05\textwidth]{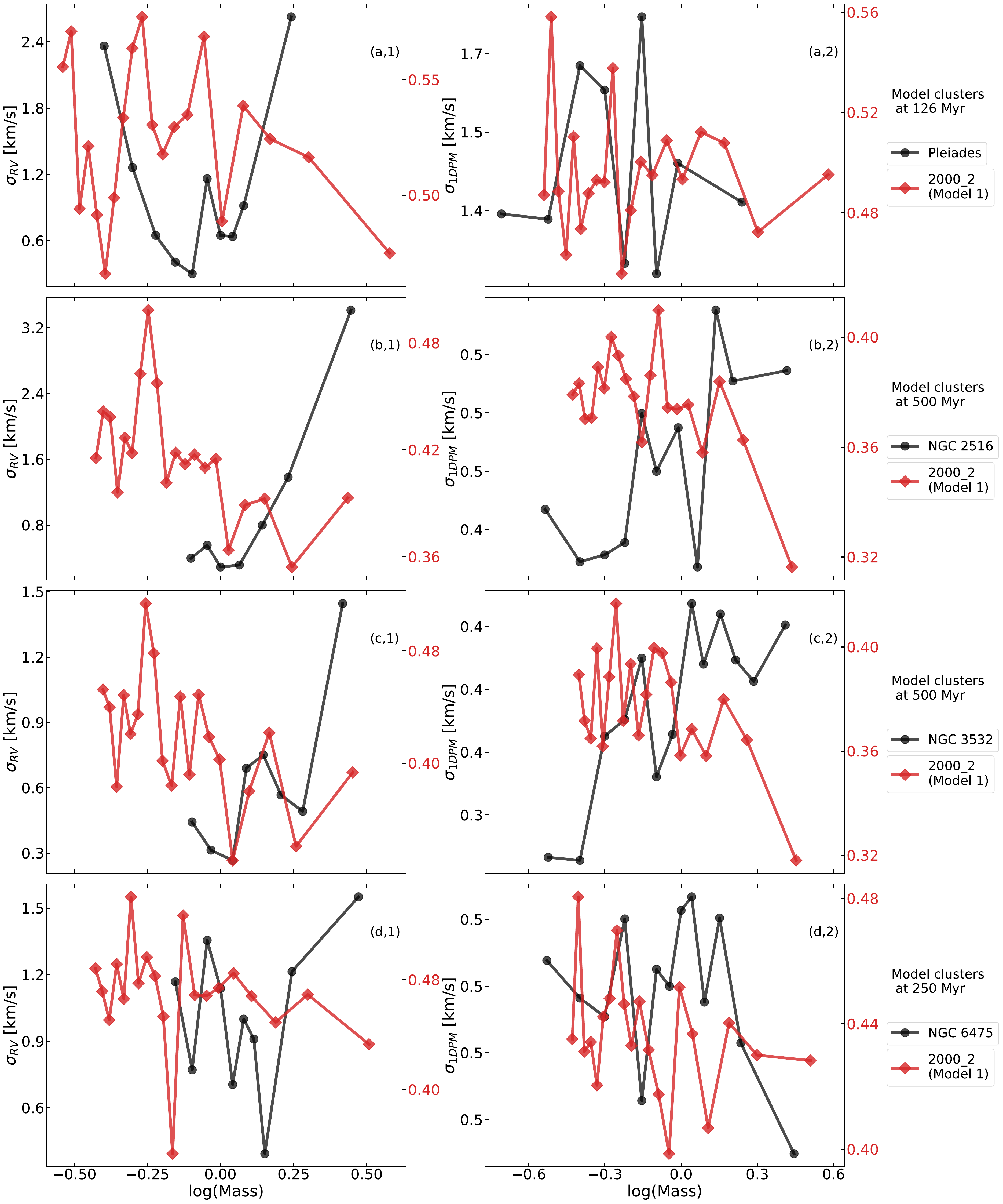}
    \caption{The PM and RV dispersion's dependence on stellar mass of four observed clusters and Model~1 ($\mcl(0)=2\times10^3~\Ms$, $\rh(0)=2~pc$) at specific evolutionary stages are presented. Only stars within tidal radius are analyzed. The black solid curves represent the 1D velocity dispersion profiles of the observed clusters (left y-axis), while the red solid curves correspond to the 1D velocity dispersion profiles of Model~1 (right y-axis). Mass from each subgroup are expressed in logarithmic scale log~(Mass).}
\label{fig:mass_model}
\end{figure}

To constrain the dynamical state of target clusters, we compare the dependence of velocity dispersions on stellar mass within the tidal radius between simulations (Model~1) and the Pleiades, NGC\,2516, NGC\,3532, and NGC\,6475 in Figure~\ref{fig:mass_model}. As the model cluster evolves, the velocity dispersion of low-mass stars increases, and that of the high-mass stars decreases. A negative dependence of velocity dispersions on stellar mass is produced. NGC\,6475's velocity dispersion profiles are most similar to those of the simulations. The cluster is likely undergoing fast expansion after the gas expulsion, which results in the observed dependence and a lower 1D PM dispersion profile compared to that of the simulations in Figure~\ref{fig:nbody0} and \ref{fig:nbody1}. However, the dependence of the velocity dispersions of NGC\,2516 and NGC\,3532 on stellar mass demonstrates an opposite trend compared to the the simulation, namely that $\sigma_{\rm RV}$ increases with higher stellar mass, implying that massive stars are being expelled from the cluster. Given that the model cluster (Model~1) host two stellar-mass BHs at the age of 250~Myr and 500\,Myr, with masses of 23~$M_\odot$ and 21~$M_\odot$, it is possible that NGC\,2516 and NGC\,3532 host more BHs that accelerate the dynamical evolution. Therefore, we propose that both NGC\,2516 and NGC\,3532 contain at least two stellar-mass BHs.

\section{Summary}\label{sec:sum}

We use Gaia DR\,3 data to analyze the dynamical states of nine open clusters: LP2373\,gp4, NGC\,1980, NGC\,2451A, NGC\,2516, NGC\,3532, NGC\,6475, NGC\,3532, UBC\,7, Praesepe and Pleiades. The stellar membership of these clusters is adopted from previous studies \citep{pang2022a,pang2022c}, where the machine learning algorithm \texttt{StarGO} is used to identify members. We further refine the cluster centroids using the \texttt{Meanshift} algorithm. The velocity dispersion of each cluster is calculated using a likelihood function based on a double Gaussian model, from which we derive the 1D velocity dispersion profiles. We also apply $N$-body simulations to generate theoretical 1D velocity dispersion profiles for clusters. By comparing the observed profiles with those of model clusters, we investigate the potential presence of stellar-mass BHs in the central regions of the nine clusters. Our findings can be summarized as follows:

\begin{enumerate}

    \item The likelihood function for proper motions and radial velocity dispersion is established using two Gaussian functions, and the broadening effects due to binary stars are simulated. Using this method, we obtain the 1D velocity dispersions ($\sigma_{\mu_{\alpha}\cos\delta}$, $\sigma_{\mu_{\delta}}$ and $\sigma_{\rm rv}$) of the 9 open clusters.

    \item To obtain the 1D velocity dispersion profiles of the nine open clusters, we divide each cluster into equal-membership subgroups based on the distances of member stars from the cluster center. The 1D velocity dispersion profiles are derived from the velocity dispersion distributions of each subgroup. Notably, a clear trend of increasing velocity dispersion with decreasing distance from the center is observed in both the proper motion and radial velocity dispersion profiles of NGC\,2516. The similar trend is also evident in the velocity dispersion profiles of NGC\,6475, NGC\,3532 and Pleiades in individual directions.

    \item We find that LP2373\,gp4, NGC\,6475 and Praesepe all exhibit a decreasing trend in the velocity dispersion with increasing stellar mass. This suggests that these clusters may be approaching energy equipartition or that they are undergoing expansion. NGC\,2516 and NGC\,3532 exhibit a velocity dispersion that increases with stellar mass. This might indicate the preferential ejection of massive stars through dynamical interactions with binary systems or BHs.

    \item Model clusters are evolved using the $N$-body code \texttt{NBODY7}. The model clusters are fitted to the 1D velocity dispersion profiles of NGC\,2516, NGC\,6475, NGC\,3532, and Pleiades. The RV dispersion profiles of these four clusters deviate from those of the simulations. Unlike the results from the simulations, the 1D PM dispersion profile of the Pleiades shows an increasing trend toward larger radii, indicating the influence of the Galactic tidal field on this cluster. For NGC\,6475, the dependence of the velocity dispersion on stellar mass agrees with the simulations, which suggests that it is undergoing significant expansion. The 1D PM dispersion profiles of NGC\,2516 and NGC\,3532 remain consistently lower than those of the model clusters. Only when evolving Model~1 to older ages (800~Myr), a better agreement is achieved with the observed velocity dispersion profiles of NGC\,2516 (123~Myr) and NGC\,3532 (398~Myr), despite their younger observed ages. We propose that this discrepancy may be attributed to the presence of at least two stellar-mass BHs in each of these two clusters.

\end{enumerate}

Our study is constrained by the data from Gaia DR\,3. The accuracy of the 1D velocity dispersion profile is significantly influenced by the completeness of the radial velocity data for cluster members. Moreover, the clusters in our sample span a limited mass range, and future studies should consider including clusters with a broader range of masses.

\begin{acknowledgments}
We wish to express our gratitude to the anonymous referee for providing comments and suggestions that helped to improve the quality of this paper.
Xiaoying Pang acknowledges the financial support of the National Natural Science Foundation of China through grants 12173029, 12573036, and 12233013. This work is supported by the China Manned Space Program
with grant No. CMS-CSST-2025-A08.  Sambaran Banerjee acknowledges funding for this work by the Deutsche Forschungsgemeinschaft
(DFG, German Research Foundation) through the project ``The dynamics of stellar-mass black holes in dense stellar systems and their role in gravitational wave generation''
(project number 405620641; PI: S. Banerjee). All $N$-body simulations are carried out on the {\tt gpudyn}-series compute servers containing NVIDIA Ampere A40 and NVIDIA RTX 2080 GPUs, located at the Argelander Institute for Astronomy, University of Bonn. The {\tt gpudyn} servers are funded by the above-mentioned DFG project and by the HISKP.

This work made use of data from the European Space Agency (ESA) mission {\it Gaia} 
(\url{https://www.cosmos.esa.int/gaia}), processed by the {\it Gaia} Data Processing 
and Analysis Consortium (DPAC, \url{https://www.cosmos.esa.int/web/gaia/dpac/consortium}).

This study also made use of the SIMBAD database and the VizieR catalogue access tool, both operated at CDS, Strasbourg, France.
\end{acknowledgments}


\software{  \texttt{Astropy} \citep{astropy2013,astropy2018,astropy2022}, 
            \texttt{SciPy} \citep{millman2011},
            and 
            \textsc{StarGO} \citep{yuan2018}.
}

\clearpage

\counterwithin{figure}{section}
\counterwithin{table}{section}

\bibliography{main}
\bibliographystyle{aasjournalv7}

\end{document}